\documentclass[iop]{emulateapj}
\usepackage{amsmath}
\usepackage{rotating}
\usepackage[colorlinks=true,citecolor=blue]{hyperref}
\shorttitle{Constraints on minimum electron Lorentz factor and matter content of jets} \shortauthors{Kang et al.}

\slugcomment{\today}  

\begin{document}
\title{Constraints on minimum electron Lorentz factor and matter content of jets for a sample of bright Fermi blazars}

\author{Shi-Ju Kang\altaffilmark{1}, Liang Chen\altaffilmark{2},
Qingwen Wu\altaffilmark{1,3}}
\altaffiltext{1}{School of Physics, Huazhong University of Science and Technology, Wuhan 430074, China} \altaffiltext{2}{Key Laboratory for Research in Galaxies and Cosmology, Shanghai Astronomical Observatory, Chinese Academy of Sciences, 80 Nandan Road, Shanghai 200030, China} \altaffiltext{3}{Corresponding author, email: qwwu@hust.edu.cn}
\begin{abstract}
We fit the (quasi-)simultaneous multi-waveband spectral energy distributions (SEDs) for a sample of low-synchrotron-peaked (LSP) blazars with a one-zone leptonic model. The seed photons that predominantly come from broad line region (BLR) and infrared (IR) molecular torus are considered respectively in external Compton process. We find that the
modeling with IR seed photons is systematically better than that with BLR photons based on a $\chi^2$ test, which suggest that $\gamma$-ray emitting region most possibly stay outside the BLR. The minimum electron Lorentz factor, $\gamma_{\rm min}$, is constrained from the modeling for these LSP blazars with good soft X-ray data (ranges from 5 to 160
with a median value of 55), which plays a key role in jet power estimation. Assuming one-to-one ratio of proton and electron, we find that the jet power for LSP blazars is systematically higher than that of FR II radio galaxies at given 151 MHz radio luminosity, $L_{\rm 151MHz}$, even though FR IIs are regarded as same as LSP blazars in unification
scheme except the jet viewing angle. The possible reason is that there are some $e^{\pm}$ pairs in the jet of these blazars. If this is the case, we find the number density of $e^{\pm}$ pairs should be several times higher than that of $e^{-}-p$ pairs by assuming the jet power is the same for LSP blazars and FR IIs at given $L_{\rm 151MHz}$.

\end{abstract}


\section{Introduction}
Blazars are most extreme and powerful sources among active galactic nuclei (AGNs), whose broadband emission is mainly dominated by non-thermal components produced in a relativistic jet pointing toward us \citep[][]{up95}. The blazars are traditionally sub-divided as flat spectrum radio quasars (FSRQs) and BL Lacertae objects (BL Lacs) according to
their emission-line features, where the BL Lacs have weak or no emission lines (e.g., equivalent width, EW, of the emission line in rest frame is less than ${\rm 5\AA}$) while FSRQs have stronger emission lines \citep[${\rm EW\geq5\AA}$, e.g.,][]{up95}. The different emission-line properties of FSRQs and BL Lacs may be triggered by accretion mode
transition, where standard cold accretion disk \citep[e.g.,][]{ss73} exists in FSRQs while advection dominated accretion flow \citep[ADAF, e.g.,][]{na94,na95} exists in BL Lacs \citep[e.g.,][]{gc01,wa02,ca03,wu06,xu09}. The spectral energy distribution (SED) of blazar generally exhibits a two-bump structure. The lower energy bump is commonly ascribed
to synchrotron radiation and peaks at infrared to X-ray band, while the second one, attributed to inverse Compton (IC) scattering, peaks at MeV-GeV band. The location of the peak for the lower energy bump in the SED, $\nu^{\rm S}_{\rm p}$, is also used to classify the sources as low (LSP, e.g., $\nu^{\rm S}_{\rm p}<10^{14}$Hz), intermediate (ISP, e.g.,
$10^{14}\rm Hz<\nu^{\rm S}_{\rm p}<10^{15}$Hz) and high-synchrotron-peaked (HSP, e.g., $\nu^{\rm S}_{\rm p}>10^{15}$Hz) blazars \citep[e.g.,][]{pg95,ab10}. In contrast to BL Lac objects, FSRQs are essentially all LSP blazars.

Origin of the seed photons for the IC scattering in blazars is an open issue, which include the synchrotron photons \citep[SSC process,][]{ma85,ma92} and/or external photons (EC process), where the external photons possibly originate from accretion disk \citep[e.g.,][]{de93}, broad line region \citep[BLR, e.g.,][]{si94,fa06}, and/or molecular torus
\citep[e.g.,][]{bl00,ar02,sm05}. Normally, the SEDs of HSP blazars (e.g., BL Lacs) seem to be consistent with the pure SSC models \citep[e.g.,][]{ma97,kr04}, while the LSP blazars often require an EC component to explain their $\gamma$-ray spectra \citep[e.g.,][]{sa99,bc02,chb11,yan14}. No EC process in BL Lacs may be caused by the disappearance of the
BLR \citep[e.g.,][]{tr01,gh02,la03,eh09,cao10} and/or torus\citep[e.g.,][]{gb07,ho08} in the low-power sources. Identifying the origin of external seed photons in FSRQs is also a diagnostic for the location of the $\gamma$-ray emitting region \citep[e.g.,][]{ag11}. For example, the external seed photons will dominantly originate from the cold disk if
the $\gamma$-ray emitting region is less than several hundreds Schwarzschild radius ($R_{\rm S}$) from the central black hole (BH), while the soft photons will mainly come from the BLR if the $\gamma$-ray emitting region is larger than several hundreds $R_{\rm S}$ and less than several thousands $R_{\rm S}$. If the $\gamma$-ray emitting region stay
outside the BLR and less than $\sim10^5 R_{\rm S}$, the most abundant seed photons are IR photons coming from the molecular torus. The cosmic background photons may be important if the gamma-ray emitting region is much larger than $\sim10^5 R_{\rm S}$ \citep[see Figures 2-3 in][]{gt09}.

The jet matter content is also a mystery, which possibly include a `normal' plasma of protons and relativistic electrons (an $e^{-}-p$ jet), a `pair' plasma consisting only of relativistic electrons and positrons (an $e^{\pm}$ jet), or combination of them. Discriminating between these possibilities is crucial in understanding the physical processes of
jet occurring close to the central engine. There are several methods were proposed to explore this issue. The first one is based on synchrotron self-absorption (SSA) arguments combined with total kinetic power of jets \citep[e.g.,][]{re96,hi05}, where the measurements on the flux and size of radio core lead to constraint on number density of
relativistic electrons and magnetic field strength in the jet. Using this method, \citet{re96} concluded that the core of M 87 is probably dominated by an $e^{\pm}$ plasma. The second one is the constraint from circular polarization in the radio core, which is due to the Faraday conversion of linear to circular polarization caused by the low-energy
electrons. \citet{wa98} suggested that the jet of 3C 279 is composed mainly by $e^{\pm}$ pairs based on its small minimum Lorentz factor of electrons ($\gamma_{\rm min}\sim1$) that derived from the circular polarizations of radio core. The third constraint comes from the cocoon dynamics of the jets. \citet{ki12} suggested that the jet of Cygnus A
should be dominated by mixture compositions of $e^{-}$-$e^{+}$-$p$ by comparing the partial pressures of electrons and protons with the observational pressure of cocoon that estimated from the interaction between jet and interstellar medium. The fourth approach is the constraint from the absence of bulk-Compton emission in FSRQs
\citep[e.g.,][]{sm00,gt10}, where the electrons in the jet traveling with a bulk Lorentz factor $\Gamma$ will interact with the photons produced by accretion disk and/or broad emission lines and will radiate at soft X-rays.  \citet{ka08} and \citet{gh12} argue against the pure $e^{\pm}$ pair jets based on the absence of observational feature of the
soft X-ray excess.

The jet power is important for understanding jet formation, disk-jet relation, jet kinetic luminosity function, and jet feedback etc. \citep[e.g.,][]{blz77,lei05,lei08,sp10,nm12,li12,wu13,cao13,cao14}. The X-ray cavities in galaxy clusters and giant elliptical galaxies provide a direct measurement of the mechanical energy released by the AGN jets
through the work done on the hot, gaseous halos surrounding them \citep[i.e.,][]{fa00,bi04,al06,ra06,bi08,ca10,wu11}. The jet power of FR II radio galaxies and radio quasars can be estimated directly from the measurements of the hotspot size and equipartition magnetic field strength along with some assumptions \citep[i.e.,][]{gs13}. \citet{ce93}
estimated the jet power from the number of emitting particles, magnetic field strength and jet velocity that constrained from the radio and X-ray data using the standard SSC theory \citep[see also][]{ta07,cg08,gu09}. Modeling the multi-waveband SEDs of blazars also returns the particle number, magnetic field strength and Doppler factor of the jet,
where the kinetic jet power can also be derived with some reasonable assumption \citep[i.e.,][]{ma03,cg08,gh09,zh14}. It should be noted that the last two methods suffer the uncertainties of the jet matter content and the low-energy cutoff, $\gamma_{\rm min}$, of electrons, where the parameter $\gamma_{\rm min}$ is coupled with the parameter of
electron number density \citep[i.e.,][]{ce93}. The value of $\gamma_{\rm min}$ can be strongly constrained in some LSP blazars (e.g., FSRQs) if their soft X-ray emission is dominated by the SSC process \citep[i.e.,][]{ta07,cg08,zh14}.

In this work, we explore the issue of possible external seed photon fields (e.g., BLR or IR torus) for a sample of LSP blazars by a $\chi^2$ test in the SED modeling. The parameter of $\gamma_{\rm min}$ can be constrained for these LSP blazars with good quality of soft X-ray data in the SED modelings. We further investigate the issue of jet power and
jet matter content if the $\gamma_{\rm min}$ is known. The sample is described in section 2. Section 3 contains the detailed information of our model. We present the results in section 4. The last section was devoted to discussion and conclusion. Throughout this work, we assume the following cosmology: $H_{0}=70\ \rm km\ s^{-1} Mpc^{-1}$,
$\Omega_{0}=0.3$ and $\Omega_{\Lambda}=0.7$.

\begin{table*}[ht]
\tiny
\caption{The parameters used to model the SED\label{tab:alldatatable}.}
\tabcolsep 0.60mm
\begin{tabular}{lllll l|llll llll|l ll}
\hline \hline
         	&	               	&	    &                    &           &        &\multicolumn{7}{c}{IR}                    	                                                                                                                            &                                                 & & \multicolumn{1}{c}{BLR}            \\
\cline{7-14}\cline{16-16}
J2000.0 name	&	   Source name  &	  $z$   &$t_{\rm var}$ 	&${\delta_{\rm var}}^{d}$&    &   $B$ (G)         & $\delta$                  & $\gamma_{\rm min}$  & ${\gamma}_{\rm b}(10^{3})$ &	   $p_{1}$              &	   $p_{2}$              &$N_{\rm 0}(10^{4})$                &$\chi^{2}$ &  &$\chi^{2}$              &     data$^e$       \\
~[1]        	&	   [2]          &	  [3]   &	   [4]  & [5]       &	&           [6]     	    &	   [7]              	&	   [8]                  &	   [9]                  &	   [10]                 &[11]   	                    &  (12)	                            &[13]       &  &       [14]             &      [15] \\
\hline
J0136+4751		&	S4	0133+47 	&	0.859	&	...   	&	20.5	&	&	1.09	$\pm$	0.24	&	27.01	$\pm$	1.93	&	8.34	$\pm$	3.94	&	0.28	$\pm$	0.06	&	1.99	 $\pm$	0.06	&	4.05	$\pm$	0.19	&	0.08	$\pm$	0.03	&	0.62	&	&	1.43	&	B	\\
J0237+2848		&	4C	28.7    	&	1.213	&	...   	&	16.0  	&	&	0.99	$\pm$	0.14	&	27.98	$\pm$	1.16	&	7.42	$\pm$	3.56	&	0.69	$\pm$	0.09	&	2.16	 $\pm$	0.04	&	4.12	$\pm$	0.15	&	0.09	$\pm$	0.02	&	0.24	&	&	0.43	&	B	\\
J0334-4008		&	PKS	0332-403	&	1.445	&	...  	&	...  	&	&	0.72	$\pm$	0.15	&	24.38	$\pm$	2.16	&	89.18	$\pm$	74.29	&	1.09	$\pm$	0.29	&	2.51	 $\pm$	0.06	&	4.42	$\pm$	0.39	&	4.78	$\pm$	1.63	&	0.52	&	&	1.44	&	B	\\
J0457-2324		&	PKS	0454-234	&	1.003	&0.78$^{a}$ &	...  	&	&	0.51	$\pm$	0.11	&	26.03	$\pm$	1.25	&$17.83^{+60.96}_{-15.83}$ 	&	0.42	$\pm$	0.10	&	2.44	 $\pm$	0.04	&	3.58	$\pm$	0.12	&	2.60	$\pm$	0.64	&	0.34	&	&	0.46	&	A	\\
J0522-3627		&	PKS	0521-36 	&	0.057	&0.43$^{a}$ &	...  	&	&	0.93	$\pm$	0.04	&	9.67	$\pm$	0.19	&	47.54	$\pm$	13.73	&	1.59	$\pm$	0.10	&	2.35	 $\pm$	0.01	&	4.51	$\pm$	0.12	&	21.05	$\pm$	1.18	&	0.03	&	&	0.23	&	B	\\
J0538-4405		&	PKS	0537-441	&	0.894	&0.53$^{b}$ &	...   	&	&	1.49	$\pm$	0.28	&	35.60	$\pm$	2.64	&	62.09	$\pm$	36.38	&	0.57	$\pm$	0.15	&	2.32	 $\pm$	0.05	&	3.83	$\pm$	0.31	&	1.22	$\pm$	0.27	&	0.50	&	&	2.22	&	A	\\
J0854+2006		&	OJ	287     	&	0.306	&1.49$^{a}$ &	16.8	&	&	1.94	$\pm$	0.28	&	14.82	$\pm$	0.81	&	63.74	$\pm$	61.21	&	0.75	$\pm$	0.14	&	2.00	 $\pm$	0.03	&	3.65	$\pm$	0.14	&	1.29	$\pm$	0.26	&	0.32	&	&	0.63	&	B	\\
J1058-0133		&	4C	1.28    	&	0.890	&	...  	&	12.1	&	&	0.95	$\pm$	0.08	&	20.44	$\pm$	0.63	&	52.61	$\pm$	35.56	&	0.52	$\pm$	0.06	&	2.22	 $\pm$	0.02	&	3.98	$\pm$	0.10	&	1.23	$\pm$	0.14	&	0.09	&	&	0.68	&	B	\\
J1130-1449		&	PKS	1127-145	&	1.184	&	...  	&	...   	&	&	1.40	$\pm$	0.13	&	18.74	$\pm$	0.66	&	84.05	$\pm$	42.16	&	0.58	$\pm$	0.07	&	2.01	 $\pm$	0.02	&	3.98	$\pm$	0.12	&	0.83	$\pm$	0.09	&	0.10	&	&	0.62	&	C	\\
J1159+2914		&	4C	29.45   	&	0.725	&0.29$^{b}$ &	28.2	&	&	0.81	$\pm$	0.19	&	32.50	$\pm$	1.97	&	79.05	$\pm$	55.27	&	0.29	$\pm$	0.08	&	2.54	 $\pm$	0.05	&	3.70	$\pm$	0.16	&	10.31	$\pm$	2.89	&	0.34	&	&	1.35	&	A	\\
J1222+0413		&	PKS	1219+04 	&	0.965	&	...   	&	...   	&	&	0.99	$\pm$	0.09	&	20.47	$\pm$	0.73	&	162.64	$\pm$	18.42	&	0.52	$\pm$	0.05	&	1.96	 $\pm$	0.02	&	4.40	$\pm$	0.14	&	0.46	$\pm$	0.04	&	0.07	&	&	0.84	&	B	\\
J1229+0203		&	3C	273     	&	0.158	&0.58$^{a}$ &	16.8	&	&	1.13	$\pm$	0.11	&	11.68	$\pm$	0.39	&	37.09	$\pm$	22.84	&	1.52	$\pm$	0.16	&	2.08	 $\pm$	0.01	&	4.15	$\pm$	0.14	&	11.20	$\pm$	0.98	&	0.07	&	&	0.16	&	A	\\
J1256-0547		&	3C	279     	&	0.533	&2.00$^{c}$ &	23.8	&	&	0.48	$\pm$	0.04	&	19.36	$\pm$	0.68	&	84.55	$\pm$	39.05	&	0.81	$\pm$	0.06	&	2.17	 $\pm$	0.02	&	5.01	$\pm$	0.13	&	0.50	$\pm$	0.06	&	0.14	&	&	0.37	&	A	\\
J1505+0326		&	PKS	1502+036	&	0.408	&	...   	&	...  	&	&	0.45	$\pm$	0.04	&	12.28	$\pm$	0.45	&	47.66	$\pm$	41.13	&	1.84	$\pm$	0.21	&	2.21	 $\pm$	0.02	&	4.31	$\pm$	0.20	&	1.28	$\pm$	0.19	&	0.14	&	&	0.19	&	A	\\
J1517-2422		&	AP	Lib     	&	0.049	&	...   	&	...   	&	&	0.50	$\pm$	0.10	&	7.72	$\pm$	0.06	&$47.25^{+118.63}_{-45.25}$ &	1.90	$\pm$	0.33	&	1.59	 $\pm$	0.04	&	4.41	$\pm$	0.36	&	0.09	$\pm$	0.02	&	0.49	&	&	0.79	&	B	\\
J1642+3948		&	3C	345     	&	0.593	&	...  	&	7.7 	&	&	0.60	$\pm$	0.09	&	21.10	$\pm$	1.18	&$21.71^{+33.72}_{-19.71}$ 	&	0.80	$\pm$	0.17	&	2.53	 $\pm$	0.03	&	4.11	$\pm$	0.21	&	5.06	$\pm$	0.99	&	0.31	&	&	1.03	&	A	\\
J1800+7828		&	S5	1803+784	&	0.680	&1.03$^{a}$ &	12.1	&	&	0.93	$\pm$	0.11	&	18.73	$\pm$	0.94	&	61.91	$\pm$	49.93	&	1.14	$\pm$	0.16	&	2.03	 $\pm$	0.03	&	4.23	$\pm$	0.20	&	0.29	$\pm$	0.05	&	0.23	&	&	0.77	&	A	\\
J1833-2103		&	PKSB 1830-210	&	2.507	&0.32$^{a}$ &	...   	&	&	0.54	$\pm$	0.20	&	53.90	$\pm$	3.73	&	4.92	$\pm$	1.82	&	0.47	$\pm$	0.11	&	2.27	 $\pm$	0.08	&	4.45	$\pm$	0.32	&	0.94	$\pm$	0.34	&	0.63	&	&	1.98	&	A	\\
J1911-2006		&	PKSB 1908-201	&	1.119	&	...   	&	...  	&	&	0.39	$\pm$	0.06	&	28.27	$\pm$	0.82	&	17.68	$\pm$	5.28	&	0.26	$\pm$	0.03	&	2.20	 $\pm$	0.03	&	4.21	$\pm$	0.08	&	0.80	$\pm$	0.12	&	0.05	&	&	0.49	&	B	\\
J1923-2104		&	PMNJ 1923-2104	&	0.874	&	...  	&	...   	&	&	1.01	$\pm$	0.10	&	20.90	$\pm$	0.84	&	56.91	$\pm$	26.48	&	1.29	$\pm$	0.19	&	2.20	 $\pm$	0.02	&	4.03	$\pm$	0.23	&	0.67	$\pm$	0.08	&	0.12	&	&	0.57	&	B	\\
J2007-4434		&	PKS	2004-447	&	0.048	&	...   	&	...  	&	&	0.45	$\pm$	0.03	&	9.05	$\pm$	0.27	&	47.78	$\pm$	21.96	&	2.65	$\pm$	0.20	&	2.26	 $\pm$	0.01	&	5.52	$\pm$	0.36	&	4.27	$\pm$	0.37	&	0.09	&	&	0.05	&	A	\\
J2148+0657		&	4C	6.69    	&	0.990	&	...  	&	15.5	&	&	0.78	$\pm$	0.11	&	18.79	$\pm$	0.83	&	12.42	$\pm$	4.53	&	0.46	$\pm$	0.08	&	2.28	 $\pm$	0.02	&	4.59	$\pm$	0.23	&	4.42	$\pm$	0.55	&	0.11	&	&	1.14	&	C	\\
J2151-3027		&	PKS	2149-307	&	2.345	&	...  	&	...   	&	&	1.10	$\pm$	0.07	&	18.03	$\pm$	0.43	&	160.98	$\pm$	13.89	&	0.99	$\pm$	0.09	&	2.24	 $\pm$	0.01	&	4.34	$\pm$	0.12	&	12.95	$\pm$	0.81	&	0.03	&	&	0.35	&	C	\\
J2202+4216		&	BL	Lac     	&	0.069	&1.06$^{a}$ &	7.2 	&	&	0.70	$\pm$	0.06	&	8.17	$\pm$	0.27	&$10.51^{+60.16}_{-8.51}$ 	&	1.77	$\pm$	0.21	&	2.22	 $\pm$	0.02	&	3.74	$\pm$	0.12	&	3.60	$\pm$	0.45	&	0.13	&	&	0.19	&	A	\\
J2203+3145		&	4C	31.63    	&	0.295	&	...  	&	6.6 	&	&	0.83	$\pm$	0.12	&	12.03	$\pm$	0.41	&	5.43	$\pm$	4.98	&	0.79	$\pm$	0.14	&	2.31	 $\pm$	0.02	&	4.44	$\pm$	0.24	&	5.70	$\pm$	0.62	&	0.07	&	&	0.38	&	C	\\
J2207-5346		&	PKS	2204-54 	&	1.206	&	...  	&	...  	&	&	1.44	$\pm$	0.23	&	18.41	$\pm$	1.04	&	123.51	$\pm$	57.21	&	0.92	$\pm$	0.18	&	2.49	 $\pm$	0.04	&	4.17	$\pm$	0.21	&	6.49	$\pm$	1.52	&	0.24	&	&	1.02	&	C	\\
J2229-0832		&	PKS	2227-08  	&	1.560	&	...  	&	15.8	&	&	0.99	$\pm$	0.13	&	24.96	$\pm$	1.31	&	111.23	$\pm$	42.56	&	0.51	$\pm$	0.09	&	2.45	 $\pm$	0.03	&	4.12	$\pm$	0.17	&	4.35	$\pm$	0.75	&	0.21	&	&	1.36	&	B	\\
J2232+1143		&	4C	11.69   	&	1.037	&0.45$^{b}$ &	15.5	&	&	1.01	$\pm$	0.22	&	24.98	$\pm$	1.11	&	88.67	$\pm$	37.43	&	0.63	$\pm$	0.12	&	2.49	 $\pm$	0.03	&	4.19	$\pm$	0.17	&	13.76	$\pm$	2.24	&	0.15	&	&	1.19	&	C	\\
J2253+1608		&	3C	454.3   	&	0.859	&	...   	&	32.9	&	&	0.85	$\pm$	0.20	&	32.98	$\pm$	2.08	&	87.89	$\pm$	51.45	&	0.42	$\pm$	0.08	&	2.07	 $\pm$	0.04	&	4.26	$\pm$	0.23	&	0.74	$\pm$	0.17	&	0.55	&	&	2.21	&	A	\\
J2327+0940		&	PKS	2325+093	&	1.843	&	...  	&	...  	&	&	0.86	$\pm$	0.11	&	25.08	$\pm$	1.21	&	165.99	$\pm$	35.98	&	0.40	$\pm$	0.05	&	2.06	 $\pm$	0.03	&	4.17	$\pm$	0.14	&	0.81	$\pm$	0.12	&	0.16	&	&	0.75	&	C	\\
\hline \hline
\end{tabular}
\footnotetext[1]{The minimum $\gamma$-ray variability timescale selected from \cite{vovk13}.} \footnotetext[2]{The minimum optical variability timescale selected from \cite{liang13}.} \footnotetext[3]{The minimum optical variability timescale selected from \cite{bott07}.} \footnotetext[4]{The Doppler factor derived from the radio variability, which is
selected from \citet{sav10}.} \footnotetext[5]{The adopted $\gamma-$ray data in the modeling, where A represent the simultaneous data, B represent quasi-simultaneous data and C represent the integrated data within 27 months.}
\end{table*}

\section{The Sample}
For purpose of this work, we select 28 LSP blazars with $\nu_{\rm p}^{\rm S}<10^{14}$ Hz from \citet{gi12}, where the broadband SEDs from radio to $\gamma$-rays are available based on (quasi-)simultaneous observations of \emph{Plank, Swift, Fermi} and some ground-based telescopes. The data from radio to X-rays for these blazars are simultaneous. For
the $\gamma$-ray data, \citet{gi12} presented the strictly simultaneous observations that accumulated during the period \emph{Plank} observation, the quasi-simultaneous observations that integrated over a period two months centered on the \emph{Plank} observation and twenty-seven months $Fermi$-LAT integration from August 4, 2008 to November 4, 2010.
The SEDs of these sources are presented in Appendix (Figures A1--6), where the simultaneous, quasi-simultaneous $\gamma$-ray data are shown with circles and squares respectively, while the triangles represent the $Fermi$ data integrated over 27 months. Two $\gamma-$ray loud LSP blazars (PKS 1502+036 and PKS 2004-447) with the quasi-simultaneous
broadband SEDs from IR to $\gamma$-rays were also selected from \cite{pal13}. In total, our sample include 30 sources. The detailed information of the sample are presented in Table \ref{tab:alldatatable}. We present LAT name and its counterpart name in columns (1) and (2) respectively. The redshift is shown in column (3). We search literatures for the
minimum variability timescales of individual objects and finally find the optical and/or $\gamma-$ray data for 11 sources \citep{bott07,vovk13,liang13}, which are reported in Column (4). To compare our modeling parameters, the Doppler factors that estimated from variability brightness temperature are also listed in column (5) for 15 sources, which are
selected from \citet{sav10}. The model parameters constrained from the SED fittings are shown in columns (6)-(14). Column (15) show the adopted $\gamma-$ray data in the modeling, where we prefer to use the simultaneous data, then use the quasi-simultaneous data if the number of simultaneous data is less than 3 and the integrated data within 27 months
will be our last choice if quasi-simultaneous data is also less than 3.

\section{The Model}
We adopt a relatively simple, one-zone, homogeneous synchrotron and inverse Compton model, which is widely used in modeling the SED of blazars \citep[e.g.,][and references therein]{gtf10}. The emitting plasma is assumed to be a spherical region with a radius of $R$. The Doppler factor
$\delta=\left[\Gamma\left(1-\beta\cos\theta\right)\right]^{-1}\approx\Gamma$ is assumed for the relativistic jet close to line of sight in blazars with a viewing angle $\theta\lesssim1/\Gamma$. The electron spectrum is described by a broken power-law distribution with a form,
    \begin{equation}
     N(\gamma )=\left\{ \begin{array}{ll}
                    N_{0}\gamma ^{-p_1}  &  \mbox{ $\gamma_{\rm min}\leq \gamma \leq \gamma_{\rm b}$} \\
            N_{0}\gamma _{\rm b}^{p_2-p_1} \gamma ^{-p_2}  &  \mbox{ $\gamma _{\rm b}<\gamma\leq\gamma_{\rm max}$,}
           \end{array}
       \right.
   \label{Ngamma}
    \end{equation}
where $\gamma_{b}$ is broken electron Lorentz factor, $p_{1}$ and $p_{2}$ represent the indices of electron distribution below and above $\gamma_{b}$. The parameter $\gamma_{\rm min}$ and $\gamma_{\rm max}$ are minimum and maximum electron Lorentz factors, $N_{0}$ is normalization of the particle distribution.

Both SSC and EC are included in our calculation, where Klein-Nishina effect is properly considered in the inverse Compton scattering \citep[see][]{ryb79, blu70}. In the EC mechanism, we consider the seed photons predominantly originate from the BLR and molecular torus respectively. For the dissipation region within the BLR (e.g., $R_{\rm diss}\lesssim
R_{\rm BLR}$), the seed photon energy density is $u_{\rm BLR}\sim f_{\rm BLR}L_{\rm d}/(4\pi cR_{\rm BLR}^2)$, where $f_{\rm BLR}\sim0.1$ is a fraction of disk luminosity, $L_{\rm d}$, that re-emitted by the broad lines. The reverberation mapping indicated that the typical size of BLR is $R_{\rm BLR}=10^{17}L_{\rm d,45}^{1/2}$ cm
\citep[e.g.,][]{kas07,ben09}, which implies the energy density of the soft photons from BLR is roughly constant with $u_{\rm BLR}=2.65\times10^{-2}$ erg cm$^{-3}$. In the jet comoving frame, $u_{\rm BLR}'=\left(17/12\right)\Gamma^{2}u_{\rm BLR}$ \citep[see][for details]{gh08,gt09}. The radiation from BLR is taken as an isotropic black-body with a peak
frequency of $2\times10^{15}\Gamma$ Hz that mainly contributed by Ly$\alpha$ line \citep[][]{gh08}.  For the case of $R_{\rm diss}>R_{\rm BLR}$, the photon field from BLR will decrease quickly \citep[e.g.,][]{gt09} and the seed photons should dominantly come from molecular torus. Similar to $u_{\rm BLR}$, $u_{\rm IR}\sim f_{\rm IR}L_{\rm d}/(4\pi
cR_{\rm IR}^2)$, where $R_{\rm IR}=2.5\times10^{18}L_{\rm d,45}^{1/2}$ cm  and $f_{\rm IR}\sim0.5$ \citep[][]{gh08}. In jet comoving frame, $u_{\rm IR}'=3\times10^{-4}\Gamma^{2}$ erg cm$^{-3}$ \citep[][]{cl07}. The radiation from the reprocessed torus is described as a black-body spectrum with a peak frequency of $\nu_{\rm IR}=3\times10^{13}$ Hz in lab
frame, which is roughly independent of the disk luminosity since $R_{\rm IR}$ scales as $L_{\rm d}^{1/2}$ \citep[][]{cl07}.

There are nine parameters $R$, $B$, $\delta$, $p_1$, $p_2$, $\gamma_{\rm min}$, $\gamma_{\rm max}$, $\gamma_{\rm b}$, $N_0$ in our model. Instead of more commonly used ``eyeball" fit, we employ a $\chi^{2}$-minimization procedure to constrain the free parameters. However, it will take too long time to get the best fit if we allow all nine parameters to
be free. In this work, we estimate the size of emitting region from the minimum variability timescale $\Delta t_{\rm var}$, which is obtained from $R=c\delta\Delta t_{\rm var}/(1+z)$, where \emph{c} is light speed and \emph{z} is redshift. The $\Delta t_{\rm var}$ of optical/$\gamma-$ray for 11 sources are collected from literatures with the average
value of $<$$\Delta t_{\rm var}$$>$$\simeq0.82$ day. For the sources with no reported minimum variability timescales, the typical value of 1 day will be adopted \citep[e.g.,][]{gh98,ab09,fos08,zh12,caow13}. The model is not sensitive to the parameter $\gamma_{\rm max}$ and we set $\gamma_{\rm max}=100\gamma_{\rm b}$ in this work which will not affect
our results. Therefore, there are seven free parameters in our SED fittings. For a given source, we generate all the parameters in a broad range, and calculate the reduced $\chi^{2}$ for these parameters. Then we derive a probability distribution of $\chi^{2}$ (e.g., $p\propto\exp(-\chi^{2})$), and the maximum probability corresponds the best-fit
parameters for this source. The $1\sigma$ uncertainty of each parameter can be derived from the Gaussian fits to the profiles of its $p$ distribution by setting other parameters at its best-fit values.

   \section{Results}
It is well known that the one-zone leptonic model cannot explain the low-frequency radio emission, which mainly originate from the large-scale jet. In this work, we consider the data with $\log \nu \geq11.5$ ($\nu\geq300$ GHz or wavelength $\lambda\leq$1 mm) in our SED modelings, where the synchrotron radiation roughly become transparent in our model
for the typical jet parameters of LSP blazars. The putative UV excess (‘big blue bump’) of four sources (J1911-2006, J2148+0657, J2203+3145 and J2232+1143) was not included in the modeling, which is thought to be produced by the standard cold accretion disk \citep[][]{ss73}. We find that these four sources tightly follow the relation between optical
luminosity and broad emission line luminosity that defined by radio quiet AGNs, while most of other LSP blazars have much brighter optical emission at given emission-line luminosity, which support that the optical emission of these four sources indeed dominantly come from accretion disk as in RQ AGNs while that of other sources should mainly come from
the jet \citep[e.g., see Figures 3 \& 4 in][and references therein]{lj06}. In Figures A1-6 of Appendix, we show the best-fit SEDs of LSP blazars with seed photons from molecular torus (left panel) and BLR (right panel) respectively. The dotted, dot-dashed, dashed and solid lines represent the synchrotron, SSC, EC and total emission respectively. We
find that the fittings with IR seed photons are systematically better than that of BLR based on their $\chi^2$ values (see Figure 1), where these $\chi^{2}$ values are shown in columns (13) and (14) in Table \ref{tab:alldatatable} respectively. There is only one source (J2007-4434) that the fitting with seed photon from BLR is better than that from the
IR torus, but the $\chi^2$ values of these two cases are still more or less similar. Therefore, our results suggest that the seed photons from molecular torus for IC should be better than that from BLR in most of the bright LSP blazars. We list the best-fit values of the parameters and their 1-$\sigma$ errors for the modeling with IR seed photons in
Table \ref{tab:alldatatable}. For example, Figure 2 shows the probability distribution of each parameter for J0522-3627, where $B=0.93\pm0.04$ G, $\delta=9.67\pm0.19$, $\gamma_{\rm min}=47.54\pm13.73$, $\gamma_{\rm b}=(1.59\pm0.11)\times10^3$, $p_1=2.35\pm0.01$, $p_2=4.51\pm0.12$ and $N_0=(2.11\pm0.12)\times10^5$. In the following analysis, we will
mainly consider the fitting results with seed photons originate from torus.

\begin{figure}
\epsscale{1.0} \plotone{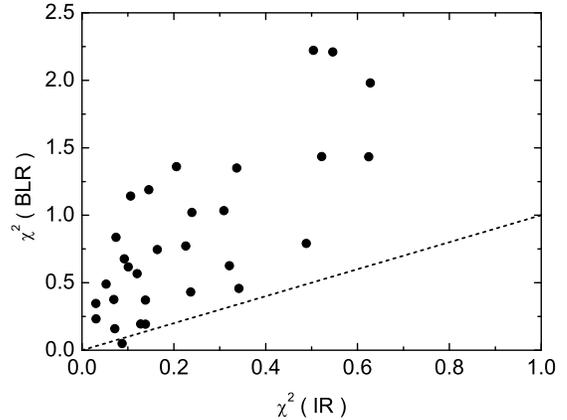} \caption{The $\chi^{2}$ values derived from multiwavelength SED fittings with the IR seed photons of torus are plotted against the fittings with the BLR seed photons. The dashed line represents $y=x$.}
\end{figure}

\begin{figure}
\epsscale{1.0} \plotone{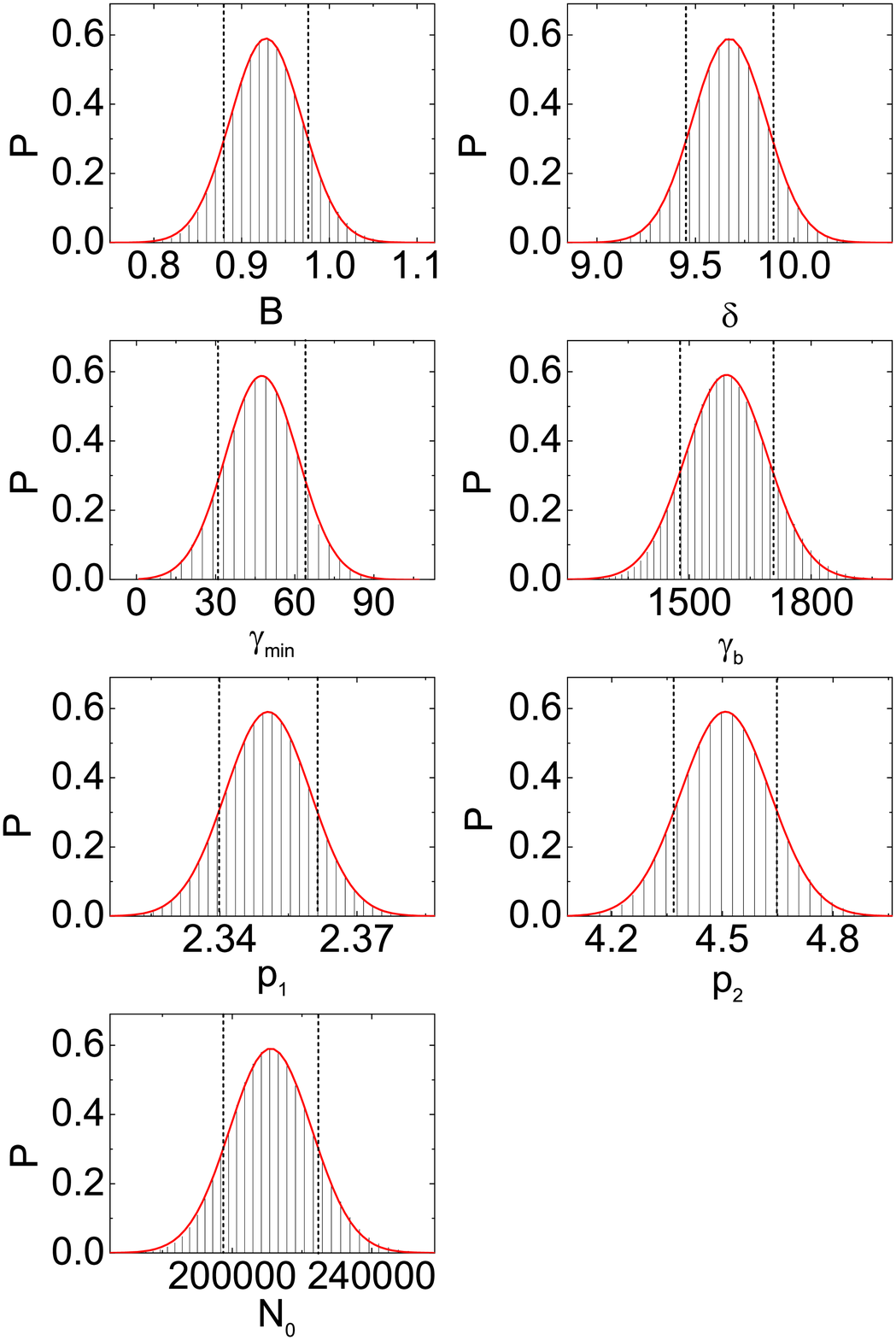} \caption{An example (J0522-3267) of probability distribution for each parameter in SED modeling along with the Gaussian fits (solid lines), where the vertical dashed lines correspond to 1$\sigma$ range of parameters.}
\label{fig:zhixinqujian}
\end{figure}

\begin{figure}
\epsscale{1.0} \plotone{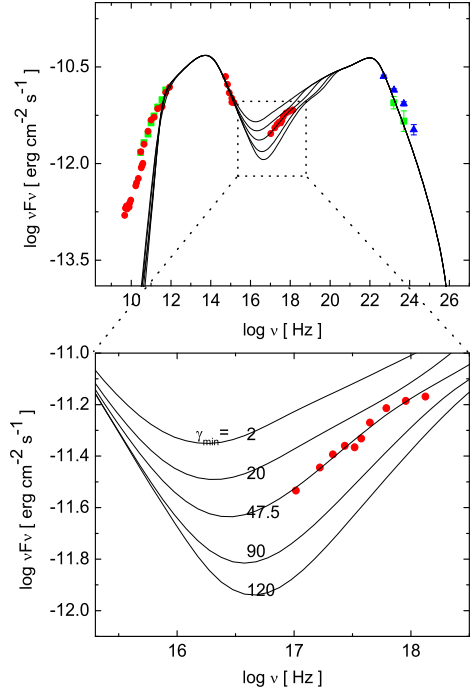} \caption{The SED of models with different $\gamma_{\rm min}$ values for J0522-3267. The solid lines from top to bottom represent $\gamma_{\rm min}=$ 2, 20, 47.5 (the best-fit value), 90, and 120 respectively.}
\label{fig:rminbianhua}
\end{figure}

\begin{figure}
\epsscale{1.0} \plotone{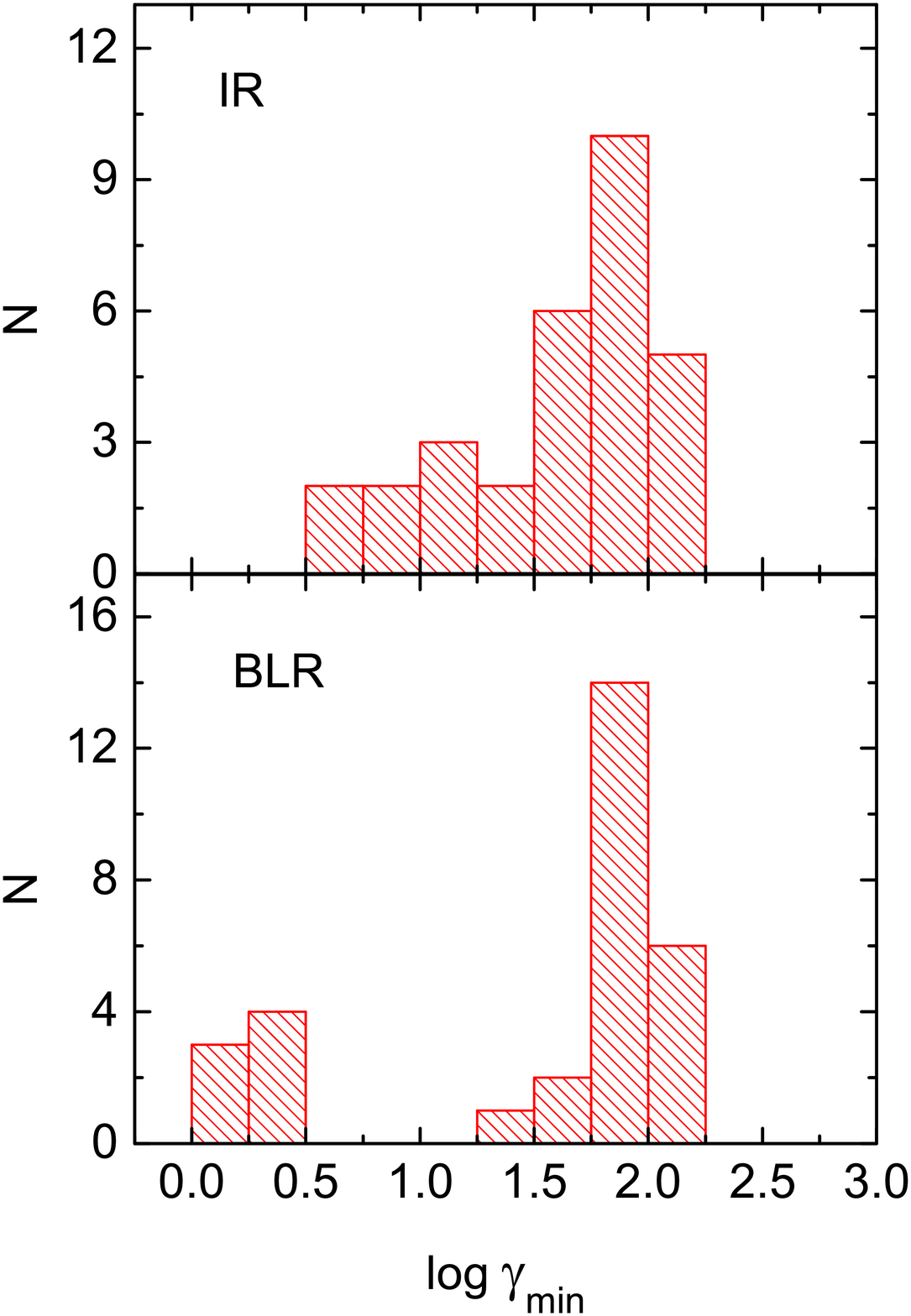} \caption{The distribution of the parameter $\gamma_{\rm min}$ for our sample, where the top and bottom panels represent the fittings with the seed photons from IR torus and BLR respectively.}
\label{fig:gammahist}
\end{figure}

In Figure 3, we present the SEDs predicted by our model with different $\gamma_{\rm min}$ values for the case of J0522-3627. It can be found that the shape of soft X-ray spectrum (e.g., 0.1--10 keV) is very sensitive to this parameter. Therefore, it is possible to constrain the parameter $\gamma_{\rm min}$ for the selected LSP blazars. Through the
fitting, we find that $\gamma_{\rm min}$ of these LSP blazars range from 5 to 160, with a median value of 55, where its distribution is shown in the top panel of Figure 4. The $\gamma_{\rm min}$ value is roughly not affected by the possible external seed photons since that the soft X-ray spectrum dominantly originate from the SSC emission. We also show
the distribution of $\gamma_{\rm min}$ that constrained from the fittings with seed photons from BLR in the bottom panel of Figure 4. It can be seen that both distributions are more or less similar (top and bottom panels).

The jet power carried by relativistic electrons, protons, magnetic field and radiation can be calculated from the parameters in our SED fittings \citep[e.g.,][]{ce93,cg08} through,
     \begin{equation}\label{pjet}
     P_{i}={\pi}R^{2}\Gamma^{2}{c}U'_{i},
    \end{equation}
   where $U'_{i}$ is the energy density of the $i$ component as measured in the comoving frame, which are given by
   \begin{equation}
   U'_{e}=m_{e}c^2\int{N(\gamma)\gamma}d\gamma,
   \end{equation}
   \begin{equation}
   U'_{p}= m_{p}c^2\int{N(\gamma)}d\gamma,
   \end{equation}
   \begin{equation}
   U'_{B}=B^2/8\pi,
   \end{equation}
   \begin{equation}
   U'_{r}=\frac{L_{\rm obs}}{4\pi R^2c\delta^4}\simeq \frac{L}{4\pi R^2c\delta^4}
   \end{equation}
   where $L_{\rm obs}$ is the total observed non-thermal luminosity and $L$ is the nonthermal luminosity derived from the modeling. Here we assume that there is one cold proton per emitting electron ($n_p=n_e$). The powers carried by each components, $P_p$, $P_e$, $P_B$ and $P_r$ are reported in Table (2).

\begin{figure}[ht]
\epsscale{1.0} \plotone{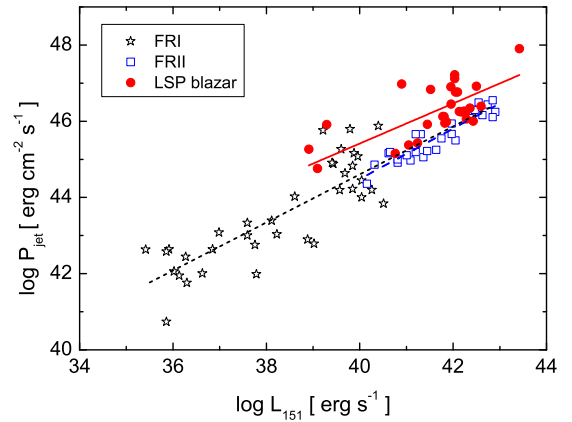} \caption{The relation between the jet kinetic power, $P_{\rm jet}$, and 151 MHz radio luminosity, $L_{\rm 151}$. The solid points represent the LSP blazars, where the jet kinetic power is derived from the parameters in the SED fitting assuming one proton per emitting electron in jet. For comparison, the empty stars and
squares represent the FR I/IIs, where the jet power is estimated from X-ray cavities (FR Is) and cocoon dynamics (FR IIs) respectively. The solid, dashed and dotted lines represent the best fits for LSP blazars, FR IIs and FR Is/IIs respectively.  } \label{fig:pjet}
\end{figure}


\begin{table*}[htbp]
\centering
\tiny
\caption{Jet power \label{tab:alldatatable2}.}
\tabcolsep 1.80mm
\begin{tabular}{lllll lll|ll}
\hline \hline
J2000.0 name	&${F_{\rm 151}}^a$	&$P_{\rm jet}^{\rm 151}$       & $P_{\rm jet}^b$        &	  $P_{B}$ 	        & $P_{e}$    	    &$P_{p}$     &$P_{r}$     &$P_{p}^{',c}$           &$\eta$  \\
            	&(Jy)    	&(${\rm erg s^{-1}}$)&(${\rm erg s^{-1}}$)  &	(${\rm erg s^{-1}}$)	&(${\rm erg s^{-1}}$)	&(${\rm erg s^{-1}}$) &(${\rm erg s^{-1}}$)	 &(${\rm erg s^{-1}}$)                 \\
~[1]        	&[2]    	&	   [3]       &[4]    	&	   [4]    	&	   [6]    	                                     &	   [7]    	&	   [8]                    & [9]                        &[10]  \\
\hline
J0136+4751	&	1.47 	&	45.51 	&	46.83 	$\pm$	0.30 	&	46.20 	$\pm$	0.20 	&	44.98 	$\pm$	0.28 	&	46.71 	$\pm$	0.32 	&	44.92 	$\pm$	0.06 	&  ...                      &  0                      \\
J0237+2848	&	2.33 	&	45.80 	&	46.77 	$\pm$	0.18 	&	46.18 	$\pm$	0.14 	&	44.85 	$\pm$	0.16 	&	46.63 	$\pm$	0.19 	&	44.98 	$\pm$	0.03 	&  ...                      & 0                      \\
J0334-4008	&	2.10 	&	45.95 	&	46.25 	$\pm$	0.32 	&	45.67 	$\pm$	0.20 	&	45.12 	$\pm$	0.32 	&	46.07 	$\pm$	0.36 	&	45.39 	$\pm$	0.07 	&	45.23 	$\pm$	0.54 	 &	0.13 	$\pm$	0.04 	\\
J0457-2324	&	2.83 	&	45.78 	&	46.90 	$\pm$	0.55 	&	45.27 	$\pm$	0.18 	&	45.30 	$\pm$	0.50 	&	46.88 	$\pm$	0.56 	&	45.12 	$\pm$	0.04 	&	45.42 	$\pm$	0.60 	 &	0.07 	$\pm$	0.03 	\\
J0522-3627	&	76.72 	&	44.87 	&	45.15 	$\pm$	0.10 	&	43.54 	$\pm$	0.05 	&	43.99 	$\pm$	0.09 	&	45.11 	$\pm$	0.11 	&	43.08 	$\pm$	0.02 	&	44.81 	$\pm$	0.11 	 &	0.67 	$\pm$	0.01 	\\
J0538-4405	&	3.67 	&	45.82 	&	46.45 	$\pm$	0.22 	&	46.22 	$\pm$	0.18 	&	44.96 	$\pm$	0.23 	&	46.03 	$\pm$	0.26 	&	45.30 	$\pm$	0.06 	&  ...                      &  0                     \\
J0854+2006	&	7.10 	&	45.23 	&	45.43 	$\pm$	0.23 	&	44.92 	$\pm$	0.15 	&	44.27 	$\pm$	0.23 	&	45.22 	$\pm$	0.26 	&	44.19 	$\pm$	0.04 	&	44.80 	$\pm$	0.31 	 &	0.55 	$\pm$	0.22 	\\
J1058-0133	&	5.56 	&	45.85 	&	46.26 	$\pm$	0.18 	&	45.59 	$\pm$	0.09 	&	45.00 	$\pm$	0.17 	&	46.12 	$\pm$	0.20 	&	44.97 	$\pm$	0.03 	&	45.30 	$\pm$	0.29 	 &	0.27 	$\pm$	0.11 	\\
J1130-1449	&	4.04 	&	45.97 	&	46.23 	$\pm$	0.15 	&	45.78 	$\pm$	0.10 	&	45.07 	$\pm$	0.15 	&	45.99 	$\pm$	0.17 	&	45.25 	$\pm$	0.03 	&	45.10 	$\pm$	0.37 	 &	0.12 	$\pm$	0.02 	\\
J1159+2914	&	4.28 	&	45.63 	&	45.95 	$\pm$	0.31 	&	45.19 	$\pm$	0.20 	&	44.77 	$\pm$	0.30 	&	45.83 	$\pm$	0.34 	&	44.66 	$\pm$	0.05 	&	45.31 	$\pm$	0.39 	 &	0.46 	$\pm$	0.17 	\\
J1222+0413	&	2.47 	&	45.72 	&	45.99 	$\pm$	0.11 	&	45.64 	$\pm$	0.10 	&	44.92 	$\pm$	0.10 	&	45.66 	$\pm$	0.11 	&	45.09 	$\pm$	0.03 	&  ...                      &  0                     \\
J1229+0203	&	97.95 	&	45.63 	&	46.13 	$\pm$	0.14 	&	44.30 	$\pm$	0.10 	&	44.98 	$\pm$	0.12 	&	46.09 	$\pm$	0.14 	&	44.85 	$\pm$	0.03 	&	45.49 	$\pm$	0.15 	 &	0.40 	$\pm$	0.01 	\\
J1256-0547	&	22.08 	&	45.99 	&	46.23 	$\pm$	0.16 	&	45.51 	$\pm$	0.09 	&	45.17 	$\pm$	0.15 	&	46.09 	$\pm$	0.17 	&	45.41 	$\pm$	0.03 	&	45.56 	$\pm$	0.21 	 &	0.45 	$\pm$	0.08 	\\
J1505+0326	&	2.49 	&	45.08 	&	45.38 	$\pm$	0.21 	&	44.06 	$\pm$	0.10 	&	44.26 	$\pm$	0.19 	&	45.32 	$\pm$	0.22 	&	43.81 	$\pm$	0.03 	&	44.98 	$\pm$	0.22 	 &	0.63 	$\pm$	0.03 	\\
J1517-2422	&	2.20 	&	43.90 	&	44.76 	$\pm$	0.20 	&	43.35 	$\pm$	0.15 	&	43.92 	$\pm$	0.18 	&	44.67 	$\pm$	0.21 	&	43.38 	$\pm$	0.01 	&	43.45 	$\pm$	0.28 	 &	0.11 	$\pm$	0.04 	\\
J1642+3948	&	12.39 	&	45.82 	&	46.77 	$\pm$	0.39 	&	45.26 	$\pm$	0.14 	&	45.22 	$\pm$	0.35 	&	46.74 	$\pm$	0.40 	&	44.94 	$\pm$	0.05 	&	45.58 	$\pm$	0.43 	 &	0.13 	$\pm$	0.03 	\\
J1800+7828	&	2.09 	&	45.47 	&	45.92 	$\pm$	0.19 	&	45.46 	$\pm$	0.13 	&	44.75 	$\pm$	0.19 	&	45.69 	$\pm$	0.22 	&	44.91 	$\pm$	0.04 	&  ...                      &  0                     \\
J1833-2103	&	10.74 	&	46.74 	&	47.90 	$\pm$	0.25 	&	45.80 	$\pm$	0.27 	&	45.87 	$\pm$	0.21 	&	47.90 	$\pm$	0.25 	&	46.00 	$\pm$	0.06 	&	46.57 	$\pm$	0.28 	 &	0.09 	$\pm$	0.01 	\\
J1911-2006	&	2.70 	&	45.91 	&	47.12 	$\pm$	0.15 	&	45.40 	$\pm$	0.14 	&	45.55 	$\pm$	0.12 	&	47.10 	$\pm$	0.15 	&	45.45 	$\pm$	0.02 	&	45.42 	$\pm$	0.24 	 &	0.04 	$\pm$	0.01 	\\
J1923-2104	&	2.73 	&	45.70 	&	46.12 	$\pm$	0.15 	&	45.69 	$\pm$	0.11 	&	44.88 	$\pm$	0.14 	&	45.89 	$\pm$	0.16 	&	45.13 	$\pm$	0.03 	&  ...                      &  0                     \\
J2007-4434	&	1.52 	&	43.83 	&	45.26 	$\pm$	0.14 	&	43.54 	$\pm$	0.08 	&	44.15 	$\pm$	0.12 	&	45.22 	$\pm$	0.14 	&	43.40 	$\pm$	0.03 	&	42.79 	$\pm$	0.38 	 &	0.01 	$\pm$	0.01 	\\
J2148+0657	&	3.50 	&	45.80 	&	47.22 	$\pm$	0.16 	&	45.28 	$\pm$	0.14 	&	45.53 	$\pm$	0.13 	&	47.21 	$\pm$	0.16 	&	45.01 	$\pm$	0.04 	&	45.50 	$\pm$	0.22 	 &	0.04 	$\pm$	0.01 	\\
J2151-3027	&	1.89 	&	46.31 	&	46.39 	$\pm$	0.08 	&	45.51 	$\pm$	0.07 	&	45.56 	$\pm$	0.06 	&	46.25 	$\pm$	0.08 	&	45.89 	$\pm$	0.02 	&	45.89 	$\pm$	0.09 	 &	0.61 	$\pm$	0.04 	\\
J2202+4216	&	1.77 	&	44.03 	&	45.92 	$\pm$	0.52 	&	43.78 	$\pm$	0.09 	&	44.26 	$\pm$	0.47 	&	45.90 	$\pm$	0.52 	&	43.49 	$\pm$	0.03 	&	43.06 	$\pm$	0.66 	 &	0.003 	$\pm$	0.002 	\\
J2203+3145	&	3.50 	&	45.01 	&	46.98 	$\pm$	0.21 	&	44.56 	$\pm$	0.13 	&	44.98 	$\pm$	0.18 	&	46.97 	$\pm$	0.21 	&	44.02 	$\pm$	0.03 	&	44.69 	$\pm$	0.28 	 &	0.01 	$\pm$	0.003 	\\
J2207-5346	&	5.65 	&	46.02 	&	46.01 	$\pm$	0.20 	&	45.78 	$\pm$	0.16 	&	44.73 	$\pm$	0.22 	&	45.56 	$\pm$	0.25 	&	44.95 	$\pm$	0.05 	&	45.50 	$\pm$	0.27 	 &	0.94 	$\pm$	0.39 	\\
J2229-0832	&	2.72 	&	46.05 	&	46.34 	$\pm$	0.19 	&	45.98 	$\pm$	0.14 	&	45.15 	$\pm$	0.19 	&	46.05 	$\pm$	0.22 	&	45.42 	$\pm$	0.04 	&  ...                      &  0                      \\
J2232+1143	&	5.66 	&	45.93 	&	46.06 	$\pm$	0.21 	&	45.31 	$\pm$	0.18 	&	44.96 	$\pm$	0.19 	&	45.93 	$\pm$	0.22 	&	44.87 	$\pm$	0.04 	&	45.71 	$\pm$	0.23 	 &	0.75 	$\pm$	0.11 	\\
J2253+1608	&	14.03 	&	46.50 	&	46.92 	$\pm$	0.25 	&	46.33 	$\pm$	0.20 	&	45.79 	$\pm$	0.22 	&	46.75 	$\pm$	0.25 	&	46.32 	$\pm$	0.05 	&  ...                      & 0                      \\
J2327+0940	&	1.41 	&	46.02 	&	46.27 	$\pm$	0.16 	&	45.87 	$\pm$	0.13 	&	45.20 	$\pm$	0.15 	&	45.98 	$\pm$	0.18 	&	45.53 	$\pm$	0.04 	&  ...                      &  0                     \\
\hline \hline
\end{tabular}
\footnotetext[1]{$F_{\rm 151}$ is the flux in 151 MHz, which is selected from NED.}
\footnotetext[2]{$P_{\rm jet}=P_p+P_e+P_B$ is the jet power in bulk motion of protons, electrons and magnetic field.}
\footnotetext[3]{$P^{'}_p$ is the jet power carried by proton after considering a fraction of $e^{\pm}$ pairs in the jet.}
\end{table*}

In Figure 5, we present the relation between the power in bulk motion of electrons, protons and magnetic field($P_{\rm jet}=P_ e+P_p+P_B$) and the 151 MHz radio luminosity $L_{151}$ for the LSP blazars, where the 151 MHz radio fluxes are selected from \emph{NED}\footnote{http://ned.ipac.caltech.edu/forms/byname.html} (see Table 2). For comparison, we
also present the $P_{\rm jet}-L_{151}$ relation for a sample of FR I (empty stars) and FR II (empty squares) radio galaxies \citep[see][for more details]{gs13}, where the jet kinetic power of FR Is and FR IIs are estimated from the X-ray cavities and cocoon dynamics respectively. We can clearly find that the jet kinetic power of blazars is
systematically higher than that of FR I/IIs at given 151 MHz radio luminosity, even though FR IIs are normally regarded as same as LSP blazars in unification scheme except their jet viewing angles. We speculate that it may be caused by the assumption of one proton per emitting electron in the jet, and the jet power will be reduced if the jet also
include a fraction of $e^{\pm}$ pairs.

To evaluate the possible positrons in the jet, we further assume $n_{p}=\eta n_{e^{-}}$ and $n_{e^{+}}=(1-\eta)n_{e^{-}}$ in the model, where $\eta=1$ corresponds to the normal jet with pure $e^{-}-p$ plasma while $\eta=0$ corresponds to the jet with pure $e^{\pm}$ pairs. In this case, the number density of protons and the jet power carried by protons
will be reduced by a factor of $\eta/(2-\eta)$, since that positrons emit with the same kind of energy spectrum as electrons. It is possible to derive the value of $\eta$ if the jet power is known from an independent method. To do this, we calculate the jet kinetic power of these LSP blazars from their 151 MHz radio luminosities using the relation of
$P_{\rm jet}-L_{151}$ for FR IIs \citep[][]{gs13},
   \begin{equation}\label{pjet1}
   P_{\rm jet}^{151}=3\pm 1\times10^{44}\left(\frac{L_{151}}{10^{25}\rm W\ Hz^{-1}\ sr^{-1}}\right)^{0.67\pm0.05},
   \end{equation}
where the jet kinetic power of FR IIs are estimated from their cocoon dynamics. $P_{\rm jet}^{151}$ derived from equation (7) for each LSP blazar is listed in Table 2. By setting $P_{\rm jet}=P^{151}_{\rm jet}$, we derive the $\eta$ values for 21 sources, which are listed in column (10) of Table 2. We noted that other nine sources have $P_B+P_e \gtrsim
P_{\rm jet}^{151}$, which may be caused by the uncertainties in jet power estimation from $P_{\rm jet}-L_{151}$ or by the uncertainties in the SED modeling. The jet power carried by protons is most possibly negligible in these nine blazars and we will simply assume $\eta\simeq 0$ in them. The distribution of $\eta$ is shown in Figure 6, where the
median $\eta$ value is 0.08 (average $\eta$ value is 0.22).
\begin{figure}[ht]
\epsscale{1.0} \plotone{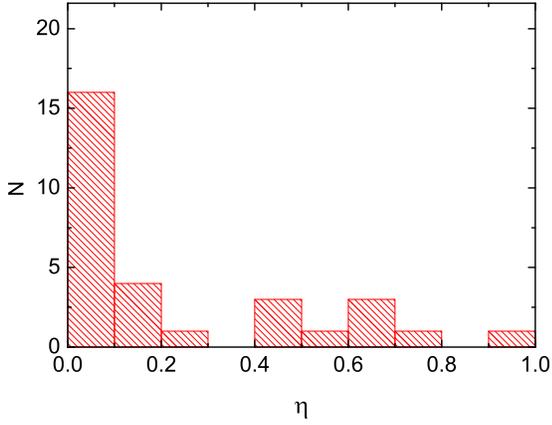} \caption{The distribution for the $\eta$ values, where $\eta=n_p/n_{e^-}$.}
\label{fig:yitahist}
\end{figure}
The jet power for cold protons, $P^{'}_p$, electrons/positrons, $P_e$, Poynting flux, $P_B$, and radiation, $P_r$, were presented in Figure 7 when considering the possible $e^{\pm}$ pairs, where the total jet power estimated from $L_{151}$ is also shown in the bottom panel (dashed lines represent median values).
\begin{figure}
\epsscale{1.0} \plotone{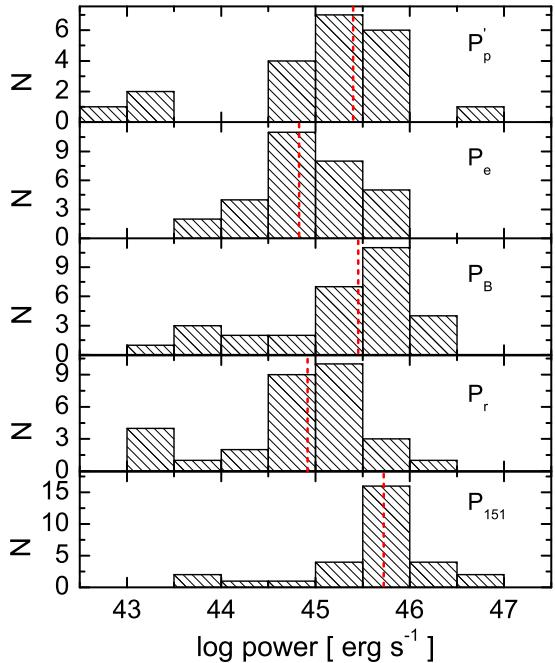} \caption{The distributions of jet power carried by cold protons ($P^{'}_p$), positrons/electrons($P_e$), Poynting flux ($P_B$), radiation ($P_r$) and total power that derived from $L_{\rm 151MHz}$ (from top to bottom panels), where we have assumed the jet include a fraction of $e^{\pm}$ pairs. The dashed line in each
panel shows its median value respectively.} \label{fig:dispjet}
\end{figure}

\begin{figure}[ht]
\epsscale{1.0} \plotone{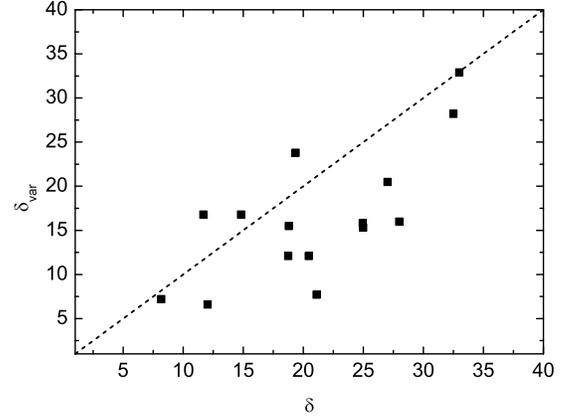} \caption{The Doppler factor derived from SED fitting versus that derived from radio variability, where the dashed line represents $y=x$. }
\label{fig:deltacompare}
\end{figure}

In Figure 8, we compare the Doppler factor, $\delta$, that derived from our SED fittings with the variability Doppler factor, $\delta_{\rm var}$, that estimated from the variability brightness temperature by assuming the intrinsic brightness temperature is limited to an equipartition value (\citealt{sav10}, see also \citealt{hova09,fanj09} for more
details). They are {positively} correlated (Pearson correlation coefficient $r=0.71$ and chance probability $p<1\%$) and roughly consistent with each other.

\section{Discussion and conclusion}
\subsection{Modeling the overall SEDs}
The simultaneous or quasi-simultaneous \emph{Plank}, \emph{Swift}, \emph{Fermi} and ground-based observations for a sample of blazars provide a great opportunity to explore their jet properties through modeling their SEDs. Our model roughly become transparent at submm waveband for the typical jet parameter of LSP blazars. The fairly good correlation of
the variability between mm and optical signals \citep[e.g.,][]{si08} or between $\gamma-$ray and mm signals \citep[e.g.,][]{we12,dam13,ori13} also support this scenario. Therefore, the emission from mm to $\gamma-$ray may originate from a more or less similar region, and it should be reasonable to fit these data with the one-zone homogeneous model.
From the fitting results, we find that the first hump from mm to optical band can be well reproduced by the synchrotron emission, while the second hump can be reproduced by SSC and EC emission for these LSP blazars (see Figures A1-6). It should be cautious that the 1$\sigma$ uncertainty of model parameter in Table (1) is derived by setting other
parameters to be their best-fit values, which may be underestimated if we allow all the parameters to be free (particularly in the case of the possible degeneracy in the model).  The observational data are not very good in a few sources (e.g., only 8 points in J1911-2006, or data is absent in a certain waveband), which will not affect our main
conclusion in a statistical sense. The consistency of the Doppler factor from the SED fitting with that derived from independent method suggest that our SED modeling should be reasonable{, even though there is no reason that the radio Doppler factor should be the same as that derived from the $\gamma$-ray emitting region \citep[e.g.,][]{fanj13}.}

\subsection{Seed photon field and location of $\gamma$-ray emitting region}

For LSP blazars, the $\gamma$-ray emission is mainly contributed by the EC process, and two main candidates of external seed fields have been proposed. The first one is the BLR if the $\gamma$-ray emitting region stay inside the BLR, which can easily explain short timescales of variability reported in some $\gamma$-ray observations of blazars
\citep[e.g.,][]{ac10,fo11}, even though the short timescales of variability do not necessarily imply the short distance to the BH. The second candidate is the molecular torus if the $\gamma$-ray emitting region stay outside the BLR and up to $\sim 10^5R_{\rm S}$ \citep[e.g.,][]{jor10,ag11}. In this work, we explore this issue through the SED modelings,
where the external seed photons from BLR or molecular torus are considered respectively. We find that the soft photons may dominantly come from IR torus in these LSP blazars based on the $\chi^2$ tests (see Figure 1), which may be caused by the KN effect. Assuming a typical seed photon frequency $\nu_{\rm ext}$, the Compton scattering by the electron
with a Lorentz factor $\gamma$ yields $\nu_{\rm EC}^{\rm T}\approx(4/3)\nu_{\rm ext}\gamma^{2}\Gamma^{2}$ within the Thomson regime in observational frame (assuming $\delta\approx\Gamma$). Combining the Thomson scattering condition $4\gamma h\nu_{\rm ext}\Gamma/m_{e}c^{2}\lesssim1$, we obtain $\nu_{\rm EC}^{\rm
T}\lesssim(1/12)(m_{e}c^{2})^{2}/(h^{2}\nu_{\rm ext})$. If seed photons come from BLR, we get $\nu_{\rm EC,BLR}^{T}\lesssim6\times10^{23}$ Hz for $\nu_{\rm ext,BLR}\approx2\times10^{15}$ Hz \citep[see also][]{caow13}. This implies that the IC emission would be significantly suppressed by KN effect if $\nu\gtrsim6\times10^{23}$Hz, which will lead to a
very steep spectrum. However, we have $\nu_{\rm EC,torus}^{T}\lesssim4\times10^{25}$ Hz $\sim165$ GeV if main seed photons come from dust torus with $\nu_{\rm ext,torus}\approx3\times10^{13}$ Hz. Therefore, the $\gamma$-ray spectrum of EC model with IR seed photons will not become very steep spectrum at observational band. This can roughly explain why
the fittings are systematically better for EC model with IR seed photons \citep[see also][]{liu06,caow13}. \citet{chb11} proposed that the IR external field may play an important role through analyzing the ratio of EC to synchrotron luminosity for a sample \emph{Fermi} bright blazars. \cite{sik09} found that the bright blazars favor the seed photons
from molecular torus over those from the BLR based on the lack of the bulk-Compton and the Klein-Nishina features in the broadband spectra.

Our modeling results give an indirect evidence that the $\gamma$-ray emitting region may locate outside the BLR and within the molecular torus, which roughly correspond to several thousand to $10^5R_{\rm S}$ from the BH. We note that the location of $\gamma$-ray emitting region should be determined by the energy dissipation or particle acceleration
processes within the relativistic jet that responsible for the generation of nonthermal electrons. \citet{asa14} explored the velocity field of M87 jet and found that the jet matter is mainly accelerated at region of $\sim 10^5R_{\rm S}$, and the jet is decelerated at larger radius. \citet{asa14}'s result suggest that most of jet energy was dissipated
at $\sim10^5R_{\rm S}$, which may correspond to the $\gamma$-ray emitting region of LSP blazars (e.g., $R_{\rm diss}>R_{\rm BLR}$). The detection of $\gamma$-rays with rest-frame energy above 20 GeV \citep[e.g.,][]{al11,pa12} suggest that the $\gamma$-ray emitting region should be located outside the highly opaque ($\tau\approx$5-10) BLR, which would
not permit photons of such high energies to escape \citep[e.g.,][]{dp03,bai09,ps10,bro13,tav13}. Based on the simultaneous flares at mm and $\gamma$-ray waveband, it was also suggested that the $\gamma$-ray emitting region should be far away from the BLR region where the high-energy photons should be produced in a region which is already transparent to
the radiation at mm wavelength \citep[e.g.,][]{jor10,ma10,ag11,dam13,ori13}. Through the correlation of mm with $\gamma$-ray light curves and direct ultrahigh-resolution 7 mm imaging with the Very Long Baseline Array (VLBA), \cite{ag11} argued that the location of $\gamma$-ray emitting region should be $>$14 pc from the BH in the jet of OJ 287, which
roughly correspond to $\gtrsim 10^4 R_{\rm S}$ if considering the BH mass $\sim 10^{10} M_{\odot}$ \citep[e.g.,][]{va12}. These independent evidences support our conclusions that the $\gamma$-ray emitting region may stay outside the BLR (several thousands $R_{\rm S}$) and the dominant seed photons for EC should be dominated by the dusty torus rather
than BLR.

\subsection{$\gamma_{\rm min}$ limits, jet power and jet composition}

It is well known that the minimum electron Lorentz factor (or the low-energy cutoff), $\gamma_{\rm min}$, play a crucial role in estimating the powers in particles (e.g., $P_{\rm jet}\propto\gamma_{\rm min}^{1-p_1}$). Normally, this parameter is poorly constrained due to its synchrotron radiation will be self-absorbed. However, these low-energy
electrons would instead contribute to the low-energy part of SSC emission. The X-ray emission of LSP blazars is known to be dominated by the SSC \citep[i.e.,][]{ta07,cg08,caow13,zh14}, which provide a possibility to constrain $\gamma_{\rm min}$ through modeling their multiwavelength SEDs (see Figure 3). In this work, we select a sample of LSP blazars
with good soft-X-ray data and find that $\gamma_{\rm min}$ ranges from 5 to 160 (with a median of 55) through their SED modelings. It is interesting to note that $\gamma_{\rm min}$ value is not sensitive to the possible external seed photon field because it was mainly constrained from the SSC process (see Figure 4). The $\gamma_{\rm min}$ values around
several tens for most of LSP blazars in our fittings are quite consistent with that reported in some of recent works even most of their fittings are evaluated by ``eyeball" \citep[e.g.,][]{tmsu00,chen10,pal13,kss13,dut13,pot13,zh14}. From a theoretical perspective, \cite{sari98} derived the minimum electron Lorentz factor $\gamma_{\rm
min}=[(p-2)/(p-1)](m_p/m_e)\varepsilon_e\Delta\Gamma$ ($p\neq2$) based on the electron energy distribution and jump conditions for a relativistic shock, where $\varepsilon_e$ is the fraction of shock energy goes into the electrons, $p$ is the index of power-law electron distribution (e.g., $N_e\propto\gamma_e^{-p}$) and  $\Delta\Gamma$ is Lorentz
factor difference for two colliding shells. We can derive $\gamma_{\rm min}\sim40$ for typical value of $p=2.24$ for relativistic shock acceleration \citep[e.g.,][]{bed98}, $\varepsilon_e=0.1$ in Gamma-ray bursts and blazars \citep[e.g.,][]{pk01,wu07} and $\Delta\Gamma\sim 1$ in blazars, which can roughly explain $\gamma_{\rm min}\sim$ several tens as
constrained from the observations. It should be noted that the lower limits of $\gamma_{\rm min}$ are not well constrained for several sources (e.g., J0457-2324, J1517-2422, J1642+3948, J2202+4216 and J2203+3145), which is caused by the poor observational data or its X-ray emission is contributed by both SSC and EC.

With the constrained of $\gamma_{\rm min}$, we calculate the jet power for each source. We find that the jet kinetic power of LSP blazars is systematically higher than that of FR IIs at given 151 MHz luminosity (see Figure 5) if assuming one proton per emitting electron in the jet, where these two types of AGNs are assumed to be intrinsically same
except the jet viewing angle \citep[e.g.,][]{up95,xu09}. One possible reason is that the jet include some positrons, which will reduce the jet power. The X-ray cavities/bubbles in galaxy clusters \citep[or giant galaxies, i.e.,][]{fa00,bi04,al06,ra06} and cocoons associated with FR I/II galaxies \citep[i.e.,][]{ki12,gs13} provide the possibility to
estimate the jet power independently. Assuming the jet power of LSP blazars is intrinsically the same as that of FR IIs at given 151 MHz luminosity, we find that the jet of most of LSP blazars should be dominated by a pair plasma, where the median ratio of proton to electron is around 0.08 or the number density of positron is around 10 times higher
than that of proton.

It should be noted that $P_{\rm jet}>P_{\rm jet}^{151}$ may also caused by some fraction of kinetic jet power will be converted into radiation before it is dissipated in hotspots on large scales. We find that the total radiative jet power normally occupy $\sim$ several percent of jet kinetic power in these LSP blazars, where most of jet radiation come
from the $\gamma-$ray emitting region. After subtracting the radiative power derived from our modeling (as an approximation for total radiative power), we find that $\eta$ value increases a little bit with a median value of 0.22 (average value is 0.29). It means that the number density of positrons is still around 4 times higher than that of protons.
The assumption of full nonthermal electrons in jet may be too simple since that only a small fraction of thermal electrons are accelerated into power-law distribution and most of electrons remains in thermal pool. If this is the case, our conclusion will be strengthened since that the radiation from these thermal electrons is much less than that of
nonthermal electrons but the jet kinetic power will increase and the $\eta$ value will decrease (i.e., more positrons are needed). It should be noted that there are still large uncertainties in estimation of jet power through $P_{\rm jet}-L_{\rm 151MHz}$ relation, and, therefore, our results should be only statistically meaningful. Based on the
possible effect of anisotropic external seed photon field in jet comoving frame, \cite{gt10} set the upper limit of $\sim10-20$ on the pair to proton number ratio \citep[see also][]{ka08}. \cite{sm00} claimed that the jets contain more $e^{\pm}$ pairs than protons based on the absence of bulk-Compton emission in FSRQs, but that jets are still
dynamically dominated by protons. Based on the VLBI observations and theory of SSA, it was also found that the jet should be dominated by $e^{\pm}$ plasma \citep[e.g.,][]{re96,hi05,du06}. Our conclusion roughly consistent with these results even they are derived from different methods.

\subsection{Conclusion}
In this work, we employ the one-zone homogeneous leptonic jet model and $\chi^2$-minimization procedure to fit the simultaneous or quasi-simultaneous multi-waveband SEDs for a sample of LSP blazars, where the external seed photons originated from IR torus and BLR are considered respectively. Our main results are summarized below.

1) The SED fitting with external seed photon from IR torus is systematically better than that from BLR. This result suggests that the $\gamma$-ray emitting region of these LSP blazars most possibly stay outside the BLR.

2) With the good quality soft X-ray data combined with other multi-wavelength observations, we find that minimum electron Lorentz factors, $\gamma_{\rm min}$, range from 5 to 160 for these LSP blazars with a median value of 55, which are not affected by the possible uncertainties of external seed photons.

3) Assuming one-to-one ratio of proton and electron in jet, we find that the jet power estimated from the fitting parameters is much higher than that of FR II galaxies at given 151 MHz radio luminosity even though they are assumed to be intrinsically the same in the unification scheme. Therefore, we propose a mixture composition of $e^{-}-e^{+}-p$ in
the jet of these LSP blazars. The number density of $e^{\pm}$ pairs should be several times higher than that of $e^{-}-p$ pairs if assuming that the jet power of LSP blazars is the same as that of FR IIs at given 151 MHz radio luminosity.

\acknowledgments We appreciate the anonymous referee for his/her helpful suggestions and comments, which help to clarify and improve our work. This work is supported by the NSFC (grants 11103003, 11133005, 11103060 and 11233006), National Basic Research Program of China (2009CB824800), and New Century Excellent Talents in University (NCET-13-0238).


\newpage
\appendix
\setcounter{figure}{0}
\renewcommand{\thefigure}{A\arabic{figure}}

\begin{figure*}[ht]
\begin{center}
\includegraphics[height=18cm,width=14cm]{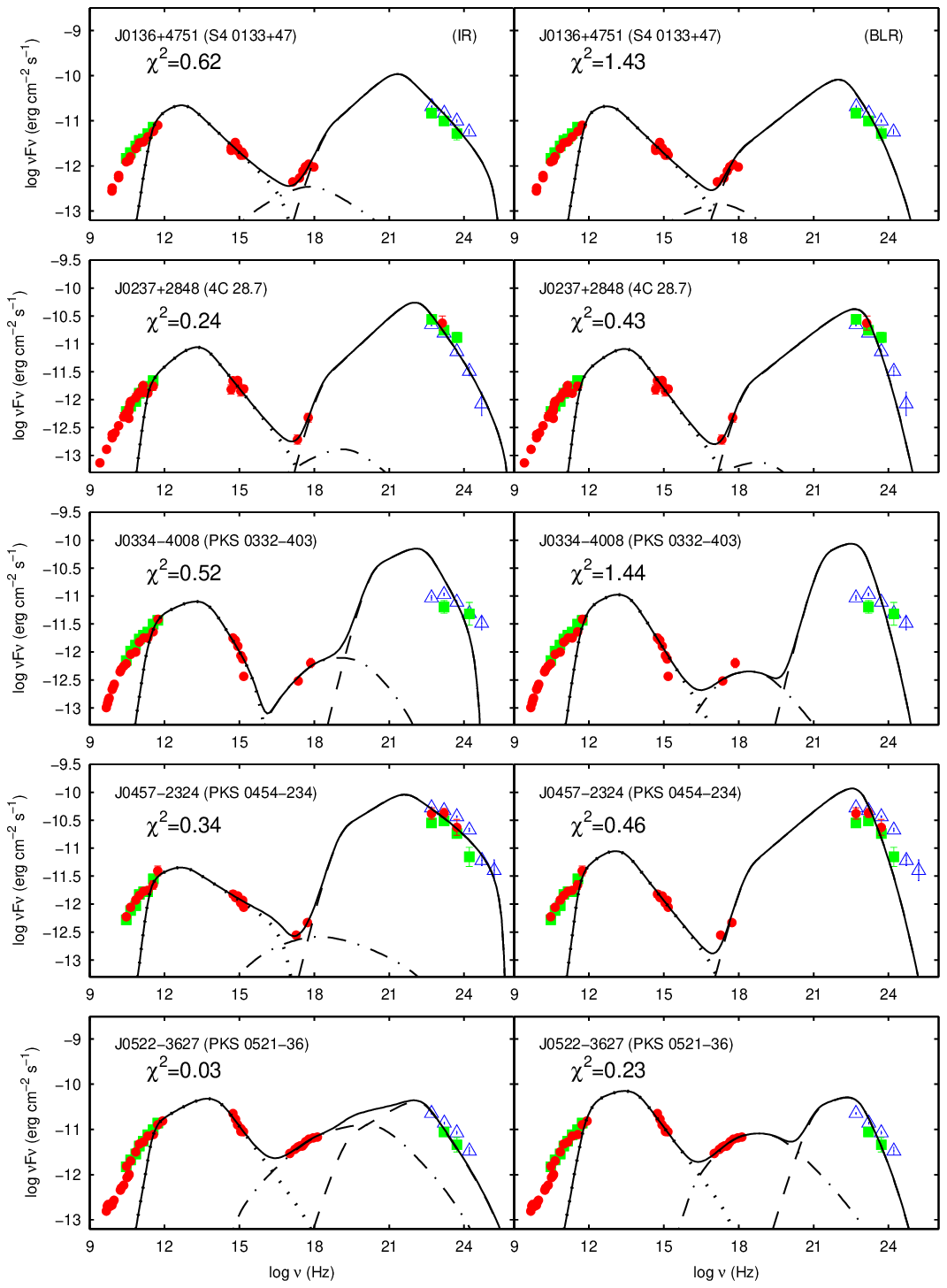}
\end{center}
\caption{SEDs of J0136+4751, J0237+2848, J0334-4008, J0457-2324 and J0522-3627, together with the fittings, where the model parameters are listed in Table \ref{tab:alldatatable}. The red circles represent the simultaneous data, the green squares represent the quasi-simultaneous data while the blue triangles represent $Fermi$ data integrated over 27
months. The left and right panels represent the fittings with seed photons from IR molecular torus and BLR respectively in the EC process. The dotted, dot-dashed and dashed lines represent the synchrotron, SSC and EC emission respectively.} \label{fig01_05}
\end{figure*}

\begin{figure*}
\begin{center}
\includegraphics[height=18cm,width=14cm]{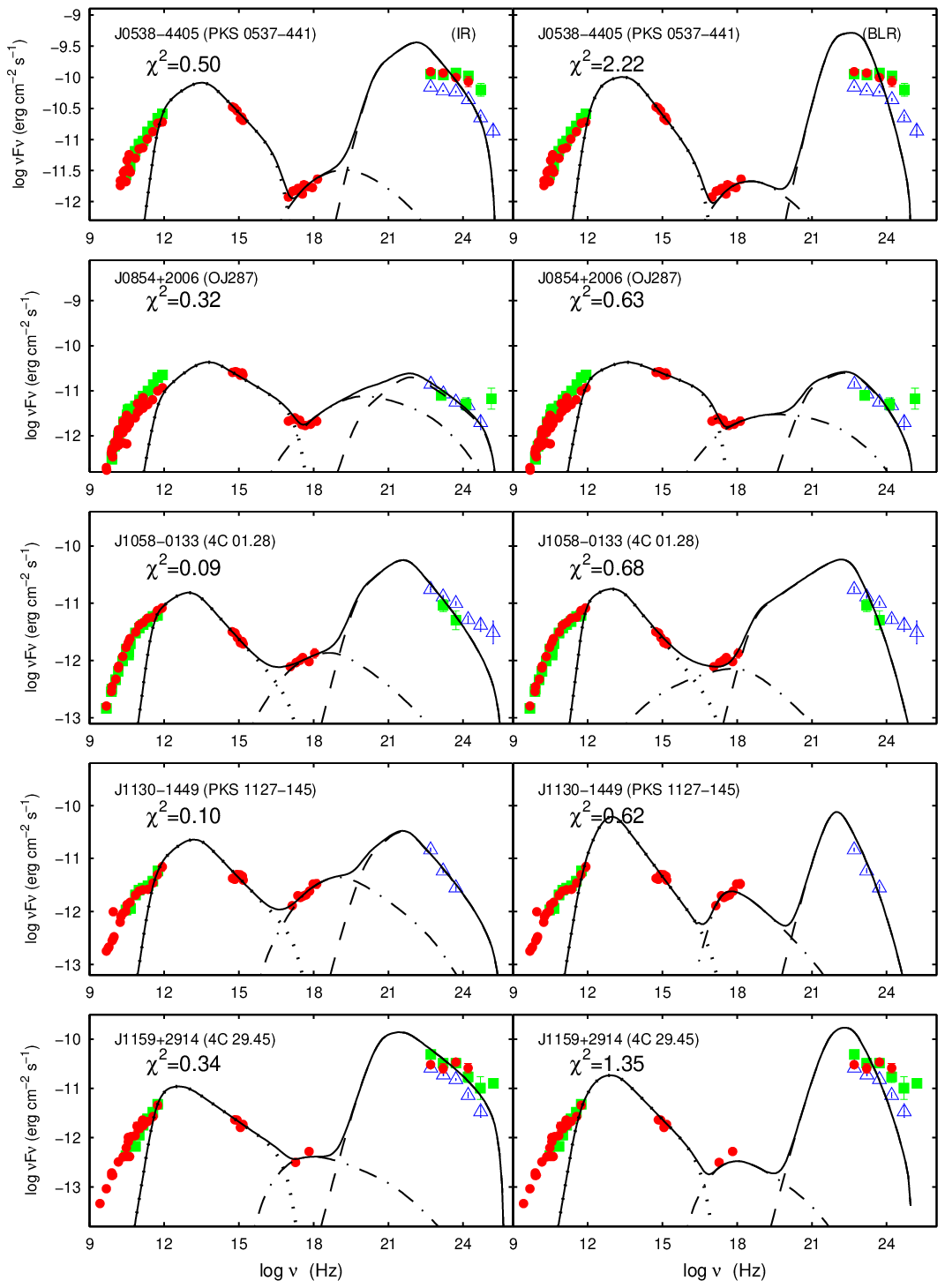}
\end{center}
\caption{SEDs of J0538-4405, J0854+2006, J1058-0133, J1130-1449, and J1159+2914. Symbols and lines as in Fig. A1. }
\label{fig06_10}
\end{figure*}

\begin{figure*}
\begin{center}
\includegraphics[height=18cm,width=14cm]{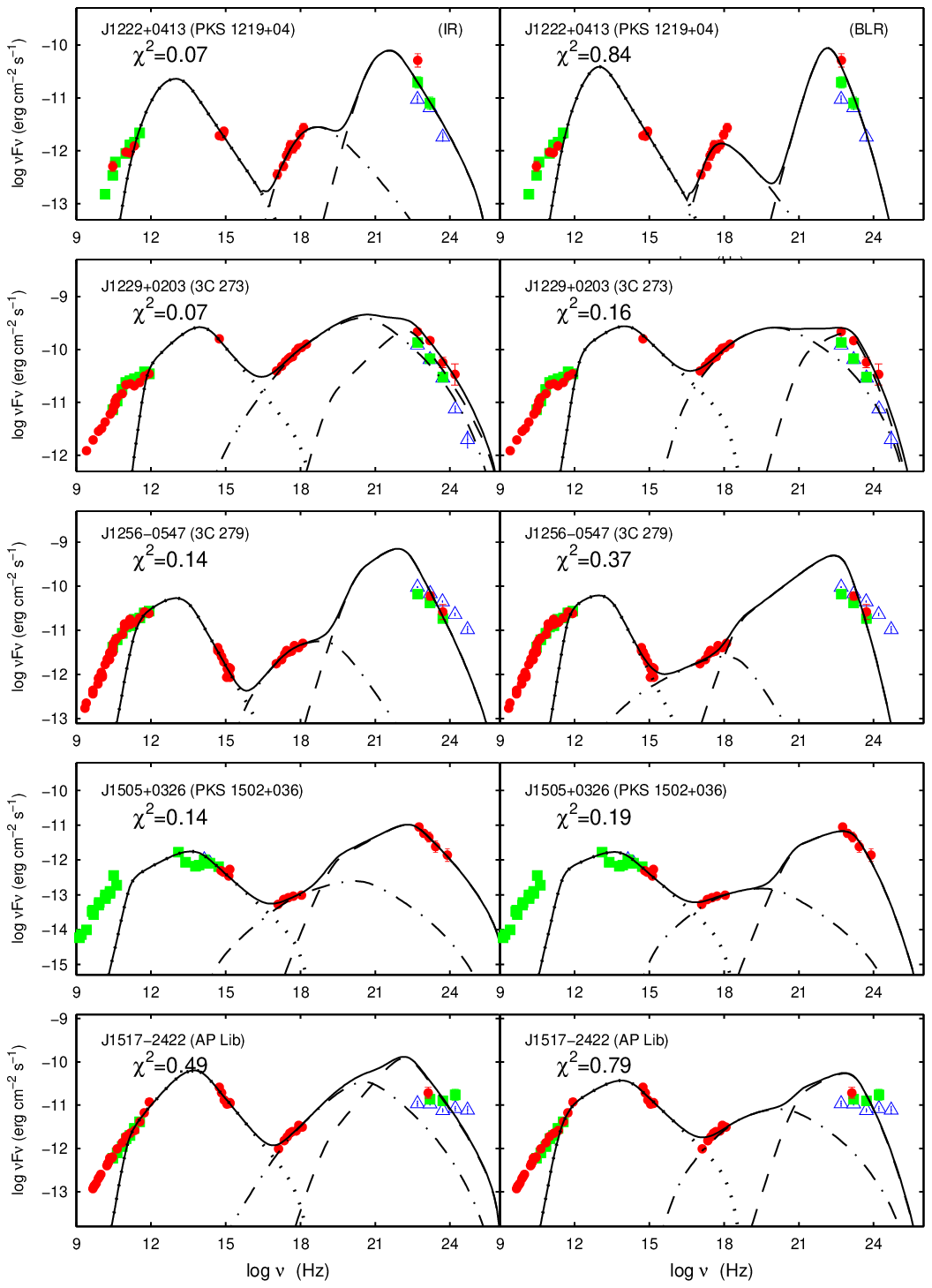}
\end{center}
\caption{SEDs of J1222+0413, J1229+0203, J1256-0547, J1505+0326 and J1517-2422. Symbols and lines as in Fig. A1. } \label{fig11_15}
\end{figure*}

\begin{figure*}
\begin{center}
\includegraphics[height=18cm,width=14cm]{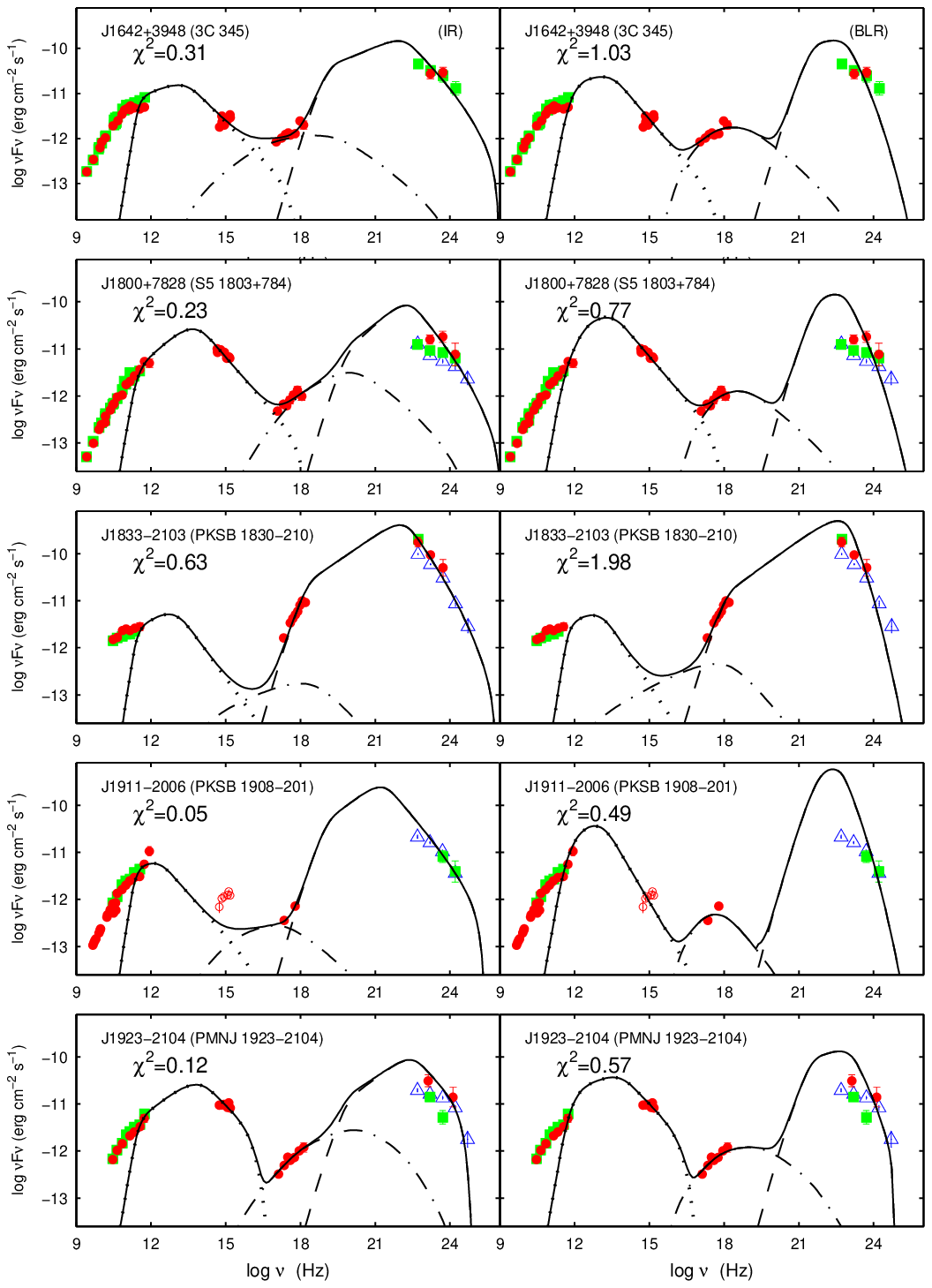}
\end{center}
\caption{SEDs of J1642+3948, J1800+7828, J1833-2103, J1911-2006 and J1923-2104. Symbols and lines as in Fig. A1. The red open circles of J1911-2006 are not included in modeling, which may originate from cold accretion disk. } \label{fig16_20}
\end{figure*}

\begin{figure*}
\begin{center}
\includegraphics[height=18cm,width=14cm]{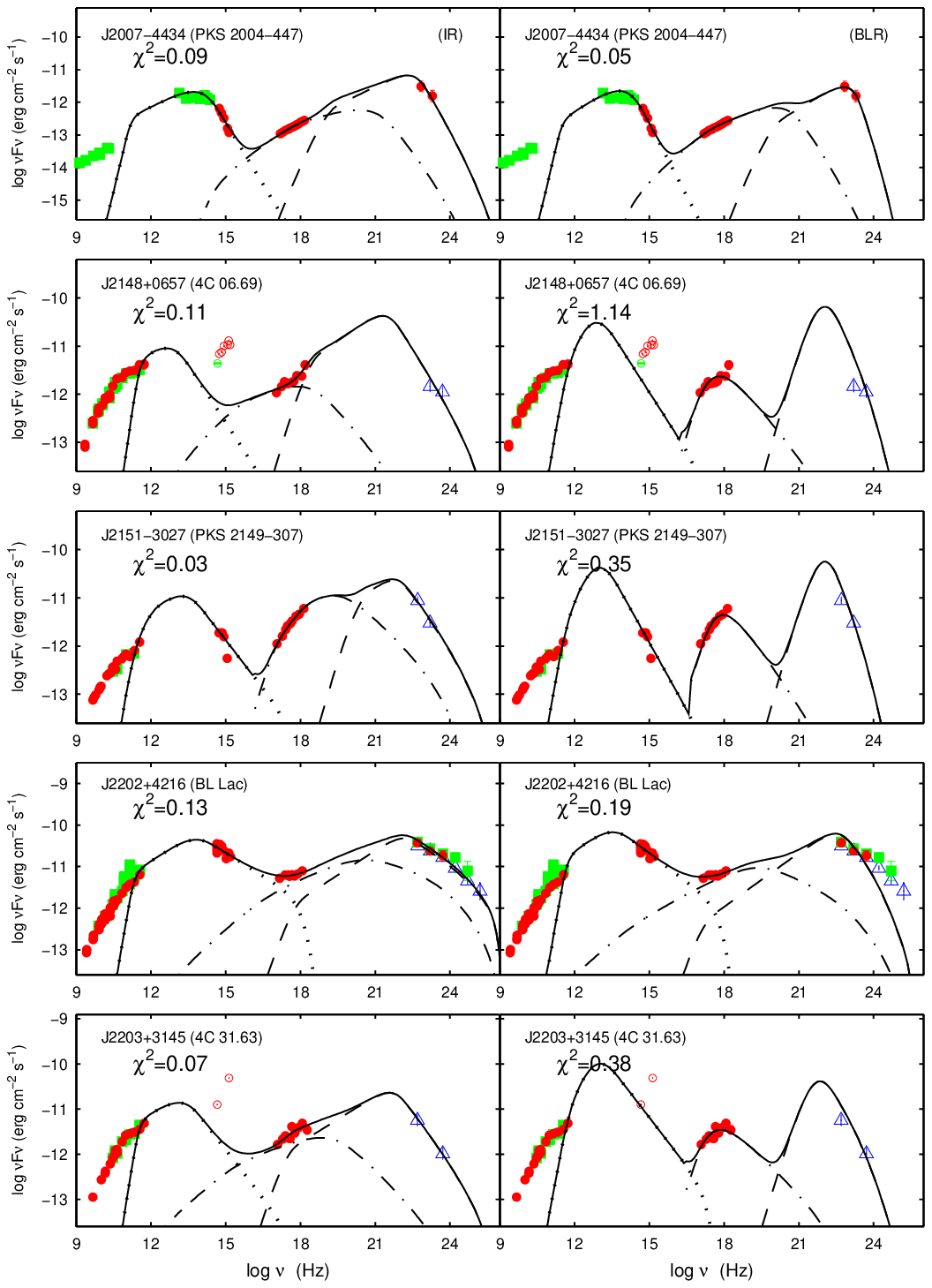}
\end{center}
\caption{SEDs of J2007-4434, J2148+0657, J2151-3027, J2202+4216 and J2203+3145. Symbols and lines as in Fig. A1. The red open circles of J2148+0657 and J2203+3145 are not included in modeling, which may originate from cold accretion disk. } \label{fig21_25}
\end{figure*}

\begin{figure*}
\begin{center}
\includegraphics[height=18cm,width=14cm]{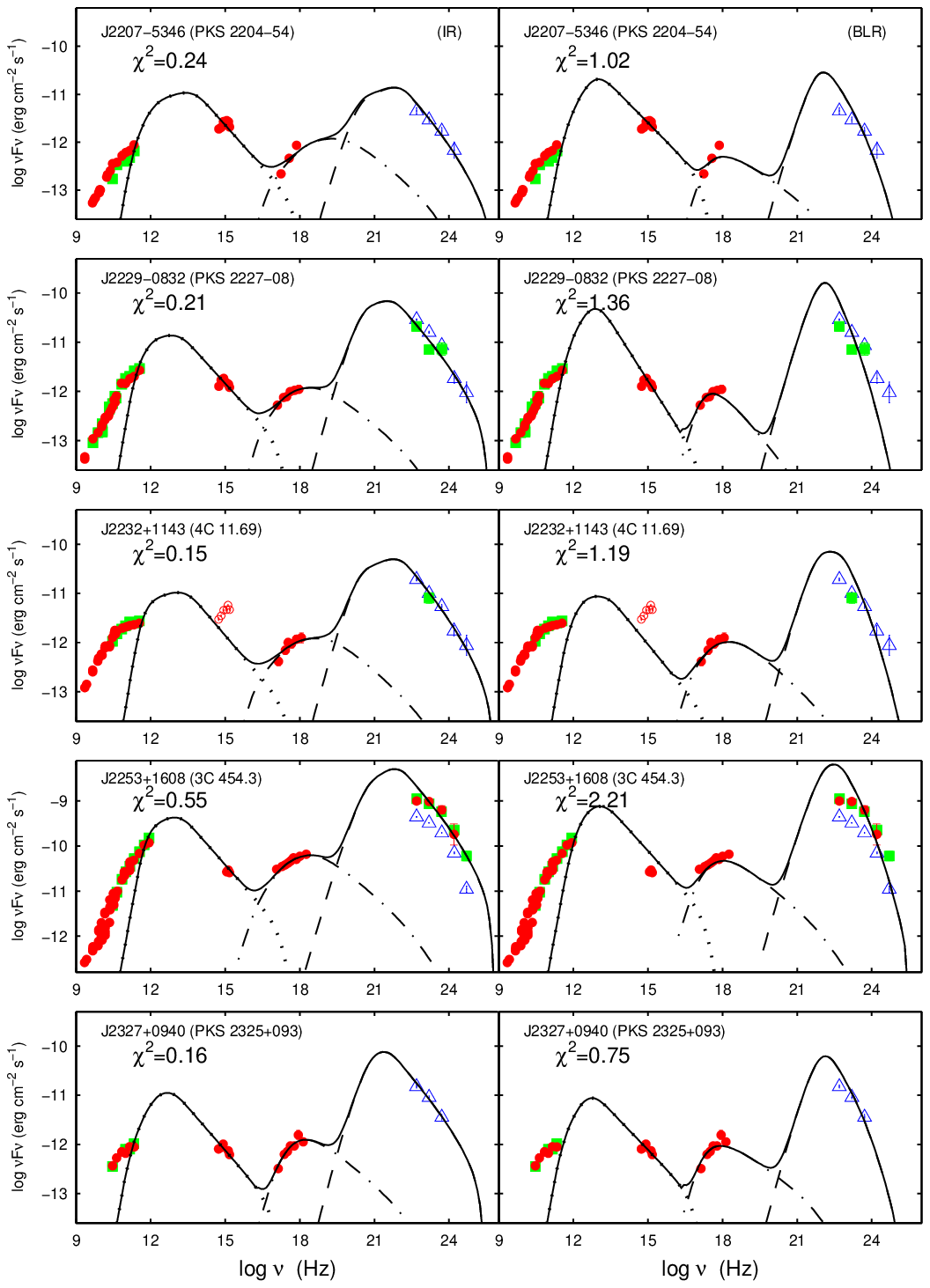}
\end{center}
\caption{SEDs of J2207-5346, J2229-0832, J2232+1143, J2253+1608 and J2327+0940. Symbols and lines as in Fig. A1. The red open circles of J2232+1143 are not included in modeling, which may originate from cold accretion disk. } \label{fig26_30}
\end{figure*}

\end{document}